\documentclass[sigconf]{acmart}

\setcopyright{none}
\acmConference[Author's final version (c) 2018 by the authors. This is the author's version of the work. It is posted here for your personal use. Not for redistribution. The definitive version was published in ACM Multimedia 2018.]
{}{}{}

\usepackage{booktabs} 

\begin{document}
\title{Few-Shot Adaptation for Multimedia Semantic Indexing}
\author{Nakamasa Inoue}
\affiliation{\institution{Tokyo Institute of Technology}}
\email{inoue@ks.cs.titech.ac.jp}
\author{Koichi Shinoda}
\affiliation{\institution{Tokyo Institute of Technology}}
\email{shinoda@c.titech.ac.jp}


\begin{abstract}
\sloppy
We propose a few-shot adaptation framework, which bridges zero-shot learning and supervised many-shot learning, for semantic indexing of image and video data.
Few-shot adaptation provides robust parameter estimation with few training examples,
by optimizing the parameters of zero-shot learning and supervised many-shot learning simultaneously.
In this method, first we build a zero-shot detector, and then update it by using the few examples.
Our experiments show the effectiveness of the proposed framework on three datasets: TRECVID Semantic Indexing 2010, 2014, and ImageNET.
On the ImageNET dataset, we show that our method outperforms recent few-shot learning methods.
On the TRECVID 2014 dataset, we achieve 15.19~\% and 35.98~\% in Mean Average Precision under the zero-shot condition and the supervised condition, respectively. To the best of our knowledge, these are the best results on this dataset.
\end{abstract}

%
%
%



\maketitle

\sloppy
\def\proc{}
\def\tablevspace{\vspace{-7mm}}
\def\tableE{
\begin{table}
\begin{center}
\begin{tabular}{|l|r|}
\hline
Methods & \scalebox{0.93}[1]{Mean AP} \\
\hline
Ours (with pseudo samples) & {\bf 35.89} \\
Ours (without pseudo samples) & 35.73 \\
\hline
Snoek \cite{Snoek_2014} (8 CNNs) & 33.19 \\
Laaksonen \cite{Laaksonen_2014} (2 CNNs + Hard Negative Mining) & 29.36 \\
Inoue \cite{Inoue_2014} (CNN + Temporal N-Gram Model) & 28.12 \\
Safadi \cite{Safadi_2015} (CNN + Re-Ranking Model) & 26.59 \\
Ballas \cite{Ballas_2014} (CNN + Audio-Visual Features) & 25.90\\
\hline
\end{tabular}
\end{center}
\caption{
\label{table:many-shot_evaluation}
{\bf Many-Shot Evaluation with and without pseudo samples.
The top 5 official submissions at TRECVID 2014 are also reported.}
}
\tablevspace
\end{table}
}

\def\tableY{
\begin{table}
\begin{center}
\begin{tabular}{|l|r|r|r|r|r|}
\multicolumn{6}{l}{(a) Novel Categories}\\
\hline
Few-Shot Methods & $N=1$ & 2 & 5 & 10 & 20\\
\hline
Hallucinating Features \cite{Hariharan_2017} & 32.8 & 46.4 & 61.7 & \underline{69.7} & {\bf \underline{73.8}}\\
Matching Network \cite{Vinyals_2016} & {\bf \underline{41.3}} & {\bf \underline{51.3}} & \underline{62.1} & 67.8 & 71.8\\
Ours & \underline{35.0} & \underline{49.7} & {\bf \underline{62.6}} & {\bf \underline{70.1}} & \underline{73.7}\\
\hline
\multicolumn{6}{l}{\vspace{-3pt}}\\
\multicolumn{6}{l}{(b) Base Categories}\\
\hline
Few-Shot Methods & $N=1$ & 2 & 5 & 10 & 20\\
\hline
Hallucinating Features \cite{Hariharan_2017} & {\bf \underline{88.4}} & {\bf \underline{87.1}} & {\bf \underline{86.6}} & {\bf \underline{85.5}} & {\bf \underline{85.0}} \\
Matching Network \cite{Vinyals_2016} & 76.7 & 77.8 & 80.6 & 82.2 & 83.3 \\
Ours & \underline{87.5} & \underline{84.8} & \underline{86.4} & \underline{85.2} & \underline{84.8} \\
\hline
\end{tabular}
\end{center}
\caption{
\label{tableY}
{\bf Analysis with novel and base categories.} (a) Top-5 accuracy for novel categories. (b) Top-5 accuracy for base categories.
}
\tablevspace
\end{table}
}

\def\tableZ{
\begin{table}
\begin{center}
\begin{tabular}{|l|r|r|r|r|r|}
\hline
Few-Shot Methods & $N=1$ & 2 & 5 & 10 & 20\\
\hline
Baseline & 43.0 & 54.3 & 67.2 & 72.8 & 75.9\\
Hallucinating Features \cite{Hariharan_2017} & 54.3 & \underline{62.1} & \underline{71.3} & \underline{75.8} & {\bf \underline{78.1}}\\
Matching Network \cite{Vinyals_2016} & \underline{55.0} & 61.5 & 69.3 & 73.4 & 76.2\\
Model Regression \cite{Wang_2016} & 46.4 & 56.7 & 66.8 & 70.4 & 72.0\\
Ours & {\bf \underline{55.2}} & {\bf \underline{63.3}} & {\bf \underline{71.8}} & {\bf \underline{76.0}} & \underline{78.0}\\
\hline
\end{tabular}
\end{center}
\caption{
\label{tableZ}
{\bf Evaluation on ImageNet dataset.} Top-5 accuracy is reported.
$N$ is the number of training examples per category.
The ResNet-10 network architecture in \cite{Hariharan_2017} is used for all experiments.
}
\tablevspace
\end{table}
}

\def\tableA{
\begin{table}
\begin{center}
\begin{tabular}{|l|l|}
\hline
Symbols & Meaning\\
\hline\hline
$c \in C$ & a concept to be detected\\
$ x \in \mathbb{R}^{d}$ & a feature vector for testing\\ 
\hline
\hline
$\mathcal{X} = \{ (x_{i}, y_{i} )\}_{i=1}^{N}$ & a set of training samples\\
$x_{i} \in \mathbb{R}^{d}$ & a feature vector for training \\
$y_{i} \in \{-1,+1\}$ & a label for training \\
$f_{\mbox{\tiny SV}} (\cdot)$ & a supervised detector \\
\hline
\hline
$\mathcal{P} = \{ g_{j} (\cdot)\}_{j=1}^{M}$ & a set of pre-trained detectors\\
$g_{j} (\cdot)$ & a pre-trained detector for $d_{j}$ \\
$d_{j} \in D$ & a concept where $D \cap C = \emptyset$\\
$\mbox{sim} (\cdot,\cdot)$ & similarity between two concepts\\
$f_{\mbox{\tiny ZS}} (\cdot)$ & a zero-shot detector \\
\hline
\end{tabular}
\end{center}
\caption{
\label{table1}
{\bf Summary of notations.}
}
\tablevspace
\end{table}
}

\def\tableB{
\begin{table*}
\begin{center}
\begin{tabular}{|l|c|c|c|c|c|}
\hline
Method & Input & \# Training Samples & \# Pre-trained Detectors & Output & Parameters\\
\hline\hline
Supervised Learning & $\mathcal{X} = \{ (x_{i}, y_{i} )\}_{i=1}^{N}$ & $N$ & $0$ & $f_{\mbox{\tiny SV}}$ & $\alpha_{i}, \gamma$\\
Zero-Shot Learning & $\mathcal{P} = \{ g_{j} (\cdot)\}_{j=1}^{M}$ & $0$ & $M$ & $f_{\mbox{\tiny ZS}}$ & $\beta_{j}, \gamma'$\\
Few-Shot Adaptation & $(\mathcal{X}, \mathcal{P})$ & $N$ & $M$ & $f_{\mbox{\tiny FS}}$ & $\alpha_{i}, \beta_{j}, \gamma''$\\
\hline
\end{tabular}
\end{center}
\caption{
\label{table2}
{\bf Summary of assumptions and parameters of each detector.}}
\tablevspace
\end{table*}
}

\def\tableC{
\begin{table}
\begin{center}
\begin{tabular}{|l|r|r|}
\hline
TRECVID Year & 2010 & 2014 \\
\hline\hline
Training video shots & 119,685 & 547,634\\
Positive in training per concept & 735 & 1,657\\
Testing video shots & 144,988 & 107,806\\
\hline
\end{tabular}
\end{center}
\caption{
\label{table:setting}
{\bf The number of video shots on TRECVID 2010 and 2014 datasets.}
Each dataset has 30 types of semantic concepts for evaluation.
}
\tablevspace
\end{table}
}

\def\tableD{
\begin{table}
\begin{center}
\begin{tabular}{|l|r|}
\hline
Zero-Shot Methods & \scalebox{0.93}[1]{Mean AP} \\
\hline
ConSE \cite{Norouzi_2014} (ImageNet-1K) & 6.39 \\
Inoue et al. \cite{Inoue_2016} (ImageNet-1K) & 8.31 \\
Ours (ImageNet-1K) & 8.49 \\
Ours (Places-365) & 11.63 \\
Ours (ImageNet-Shuffle13K) & 14.89 \\
Ours (3-Net Fusion) & {\bf 15.19} \\
\hline\hline
Webly Supervised Methods & \scalebox{0.93}[1]{Mean AP} \\
\hline
Jiang et al.\cite{Jiang_2014b} & 1.21\\
McGuinness et al. \cite{McGuinness_2014} & 7.97\\
\hline
\end{tabular}
\end{center}
\caption{
\label{table:zero-shot_evaluation}
{\bf Zero-Shot Evaluation.}
Our proposed method is compared with zero-shot learning methods and webly supervised learning methods.
}
\tablevspace
\end{table}
}

\def\figw#1#2#3{
\begin{figure*}[t]
\begin{center}
\includegraphics[width=#2\linewidth]{#1g.eps}
\caption{#3}
\label{#1}
\end{center}
\vspace{-2mm}
\end{figure*}
}

\def\figv#1#2#3{
\begin{figure*}[t]
\begin{center}
\includegraphics[width=#2\linewidth]{#1g.eps}
\caption{#3}
\label{#1}
\end{center}
\vspace{-2mm}
\end{figure*}
}

\def\fig#1#2#3{
\begin{figure}[t]
\begin{center}
\includegraphics[width=#2\linewidth]{#1g.eps}
\caption{#3}
\label{#1}
\end{center}
\vspace{-2mm}
\end{figure}
}

\section{Introduction}
\label{sec:intro}

With advances in information technologies, the amount of multimedia data such as video, image, audio, and text data has been increasing rapidly.
Detecting semantic concepts is known to be a fundamental technology to improve the performance of many multimedia applications including search \cite{Snoek_2010, Awad_2016}, summarization \cite{Zhao_2017, Mei_2013}, and surveillance \cite{Xian_2017b, Zhang_2014}.
Here, semantic concepts are objects, actions, and scenes.

How to bridge the semantic gap \cite{Smeulders_2000}, the lack of correspondence between low-level features and high-level semantic concepts?
This is the most important problem to be solved in semantic concept detection.
Previous studies have proved that supervised learning with many examples, i.e., supervised many-shot learning \footnote{hereinafter it is simply referred to as supervised learning}, is a straightforward way to find a mapping from low-level features to high-level semantics.
For example, support vector machines (SVMs) \cite{Drucker_1997,Vapnik_1995} and deep neural networks \cite{Krizhevsky_2012, Simonyan_2015, Szegedy_2015} have been shown to be effective in video semantic indexing
\cite{Awad_2016, Snoek_2014} and object recognition
\cite{Krizhevsky_2012, Simonyan_2015, Szegedy_2015, Perronnin_2010}.
These methods require large-scale training data in which positive and negative labels of semantic concepts for each image/video are given.
However, the cost of collecting training data increases as the number of target semantic concepts increases, since manual annotation is needed.

To reduce such costs, some researchers are focusing on
techniques to train statistical models with few training samples.
For example, few-shot learning in \cite{Hariharan_2017} has introduced
artificially generated features, namely hallucinating features,
which can be used as additional samples for network training with few images.
Matching networks \cite{Vinyals_2016} for one-shot learning
has provided a framework to adapt an image embedding network to a given training image with attention mechanism.
Domain adaptation \cite{Oquab_2014,Babenko_2014} and Bayesian estimation \cite{Mensink_2013, Perronnin_Dance_Csurka_Bressan_2006} are also known to be effective.
However, these methods rely on the assumption that given few training samples are high-quality, i.e., an object is often at the center of an image without occlusion or noise.
Thus, they are not always effective for concept detection from images/videos in the wild with low-quality training samples.


\fig{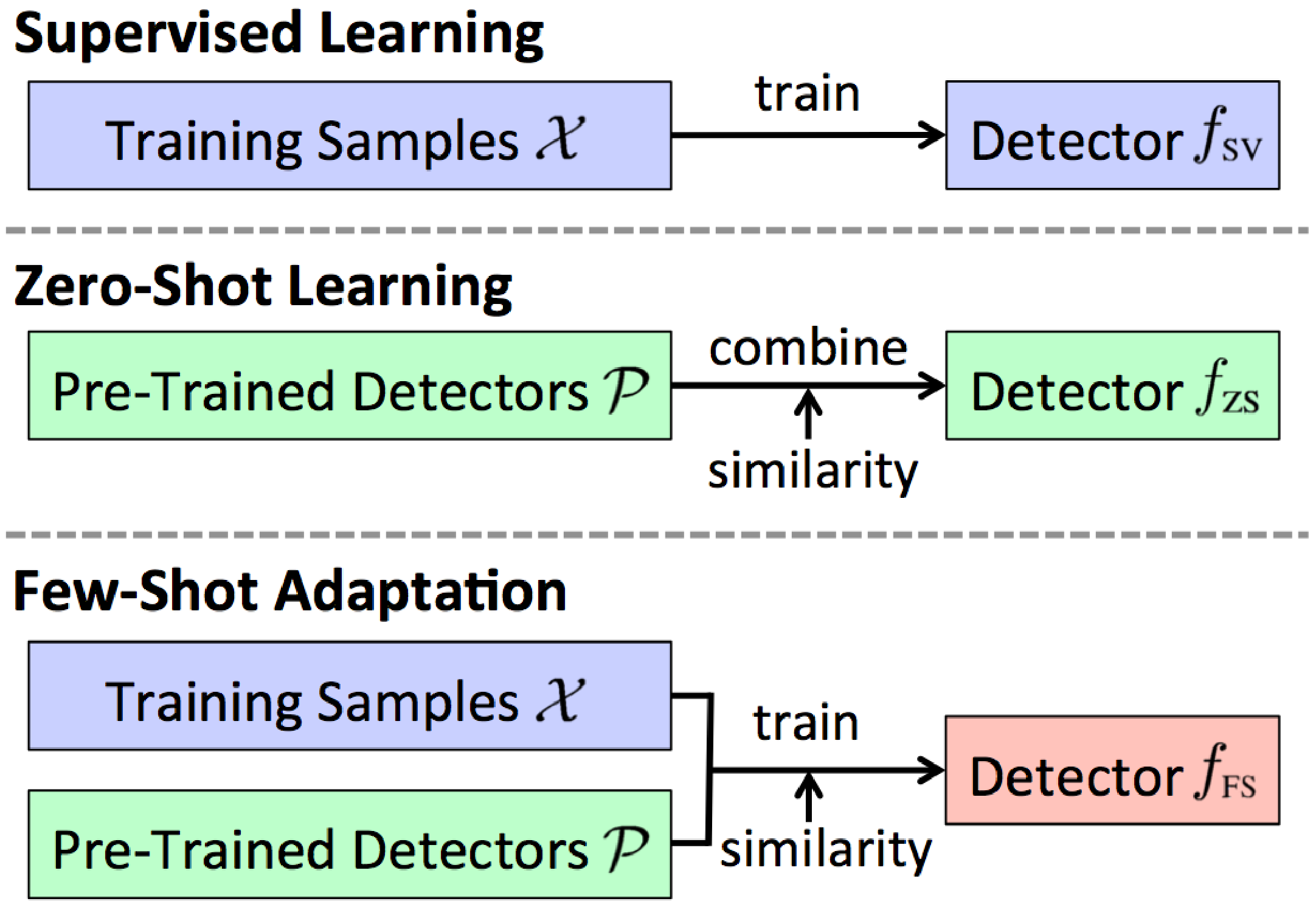}{0.95}{
{\bf Few-Shot Adaptation Framework.}
Few-shot adaptation combines supervised many-shot learning and zero-shot learning.
To train a detector, few-shot adaptation accepts two inputs: a set of training samples $\mathcal{X}$ and a set of pre-trained detectors $\mathcal{P}$ from supervised and zero-shot learning frameworks, respectively.
}

Another effective approach, which does not rely on the high quality of images,
is to utilize semantic relation among concepts obtained from large-scale text data.
Recent zero-shot learning studies \cite{Norouzi_2014, Frome_2013, Mensink_2014, Jain_2015, Xian_2017, Cappallo_2017} have shown that combining pre-trained detectors based on semantic relation is effective.
For example, convex combination of pre-trained detectors for 1,000 objects has been proposed
to make effective detectors for other unseen objects \cite{Norouzi_2014, Frome_2013} or actions \cite{Jain_2015, Qin_2017, Xu_2016, Gan_2016}.
In these methods, word vectors, which represent a word by a real-valued vector, e.g., {\it word2vec} \cite{Mikolov_2013, Mikolov_2013_2},
are often introduced to measure similarity among objects and/or actions, and are used to determine weights for the convex combination.
Some recent studies have focused on applications of zero-shot learning to the other learning frameworks.
For example, prior knowledge from zero-shot learning is introduced to active learning in \cite{Gavves_2015}.
In few-shot learning, we believe that techniques for zero-shot learning and supervised learning benefit from each other, because their inputs are different and complementary.

In this paper, we propose a few-shot adaptation framework, which bridges supervised learning and zero-shot learning for image and video semantic indexing.
It optimizes the parameters of supervised learning and zero-shot learning, simultaneously,
under an assumption that a set of training samples and a set of pre-trained detectors are given (Figure~\ref{fig001}).
In our experiments, the proposed framework is evaluated on three datasets: TRECVID Semantic Indexing 2010, 2014, and ImageNet.
We achieve 15.19 \% and 35.98 \% in Mean Average Precision under the zero-shot condition and the official supervised condition, respectively.
To the best of our knowledge, these are the best results on the TRECVID 2014 dataset.

The rest of this paper is organized as follows.
Section \ref{sec:relatedwork} summarizes related studies.
Section \ref{sec3} defines the notations for supervised learning and zeros-shot learning for preparation.
Section \ref{sec:proposedmethod} presents the proposed few-shot adaptation framework.
Section \ref{sec:experiments} reports the results of experimental evaluations,
and Section \ref{sec:conclusion} describes conclusion and future work.

\section{Related Work}
\label{sec:relatedwork}
\subsection{Supervised Learning and Adaptation}
Supervised learning is a straightforward way to obtain detectors of semantic concepts from training samples.
It requires positive and negative samples for training.
Recent studies have shown the effectiveness of deep learning using convolutional neural networks (CNNs) on large-scale datasets.
AlexNet \cite{Krizhevsky_2012} with 8 layers is their typical example.
It accepts raw image data as input to train object classifiers at the final softmax layer.
GoogLeNet \cite{Szegedy_2015}, VGGNet \cite{Simonyan_2015}, ResNet \cite{He_2016}, and DenseNet \cite{Huang_2017} are its extension to deeper networks.
These networks are often trained on a large-scale image dataset such as
the ImageNet Large-Scale Visual Recognition Challenge (ILSVRC) dataset \cite{Russakovsky_2015} and the Places 365 dataset \cite{Zhou_2017}.

Some recent studies have focused on techniques to train statistical models with a small number of training samples.
Examples include one-/few-shot learning \cite{Hariharan_2017, Vinyals_2016, Santoro_2016, Kwitt_2016},
model transformation \cite{Wang_2016},
domain adaptation \cite{Oquab_2014,Babenko_2014}, and
Bayesian estimation \cite{Mensink_2013, Perronnin_Dance_Csurka_Bressan_2006}.

Hariharan {\it et al.} \cite{Hariharan_2017} proposed a few-shot learning method,
which uses artificially generated features, called hallucinating features, for additional training data.
They introduced a network to predict and generate features based on analogies on the ImageNet dataset.
For example, from a given image of a bird with a sky background,
it predicts features of bird images with other backgrounds such as forest,
by using images of other objects with these backgrounds in the ImageNet dataset.
Matching networks in \cite{Vinyals_2016} tackled a one-shot learning problem
by introducing attention mechanism to adapt an image embedding network to given one example.

For generative models such as Gaussian Mixture Models (GMM),
Bayesian estimation \cite{Mensink_2013, Perronnin_Dance_Csurka_Bressan_2006, Inoue_Shinoda_2012} is known to be effective.
Perronnin {\it el al.} \cite{Perronnin_Dance_Csurka_Bressan_2006} proposed vocaburary adaptation using GMMs.
It is extended to image representation called Fisher vectors \cite{Perronnin_Jorge_Mensink_2010}.
Mensink {\it el al.} \cite{Mensink_2013} applied maximum a posterior estimation to metric learning for image classification.
They showed the effectiveness of distance-based classifiers in image recognition with a small set of training samples.
For recent network-based discriminative models such as the above CNNs,
fine-tuning is the most promising approach.
It is effective for many tasks including event detection \cite{Mettes_2016, Habibian_2016} and action recognition \cite{Feichtenhofer_2016, Sun_2015}.

For video semantic indexing, updating or replacing only the final classification layer is one of the best ways in fine-tuning.
For example, the final softmax layer is replaced by support vector machines (SVMs) to detect semantic concepts from video data in \cite{Snoek_2015, Mettes_2016}.
This performed the best at the semantic indexing competition in the TRECVID workshop \cite{Snoek_2015, 2015trecvidover, Jiang_2014b, McGuinness_2014}.

However, with few training examples, these methods are often sensitive to the selection of training examples.
Reducing the sensitivity is known to be one of the challenging tasks in training using few examples.

\subsection{Zero-Shot Learning}
Zero-shot learning \cite{Norouzi_2014, Lampert_2013, Xian_2017}, for the case when a set of semantic concepts for training and that for testing are disjoint, has been receiving attention in recent years.
Compared with supervised learning, the performance of zero-shot learning is rather low, because image/video samples are not given for training.
A recent trend is to build detectors by a weighted combination of pre-trained detectors,
in which weights are determined based on similarity between concepts \cite{Changpinyo_2016, Xian_2017, Qin_2017, Xu_2016, Jain_2015}.
Here, how to use the relation among concepts to measure the similarity is a key factor to improve the performance.

For object recognition, the object attributes are useful to measure similarity between objects.
For example in animal recognition, attributes such as color, texture, and shape are known to be useful for zero-shot learning \cite{Lampert_2013, Huang_2015, Yu_2010}.
The more detailed the attributes are,
the more precise similarity between concepts will be.
However, it is often given to a small set of objects,
and is difficult to manually prepare the attributes for a large number of objects.

Another approach to measure the similarity between concepts
is to utilize word vectors, which represent a word by a real-valued vector, obtained from large-scale text data.
For example, word vectors extracted by the skip-gram model \cite{Mikolov_2013_2}
are introduced to zero-shot object recognition \cite{Changpinyo_2016}.
Here, the skip-gram model is a neural network to extract vector representation of words.
It is often trained on a large-scale text corpus such as Wikipedia \cite{Mikolov_2013_2}.
Since words in a text corpus include not only nouns but also verbs,
these methods have wide-range applications related to actions \cite{Qin_2017, Xu_2016, Jain_2015}, events \cite{Yu_2012, Jiang_2014, Habibian_2014, Wu_Bondugula_Luisier_Zhuang_Natarajan_2014}, and semantic concepts \cite{Inoue_2016}.
Some recent studies focus on improving word embedding methods with joint learning \cite{Zhang_2016, Xian_2016, Zhang_2017, Kodirov_2017}.

Compared with supervised learning, zero-shot learning effectively introduces knowledge from text data.
This is the reason why we believe that supervised learning and zero-shot learning benefit from each other.

\section{Supervised Learning and Zero-Shot Learning}
\label{sec3}
This section briefly summarizes the supervised and zero-shot learning methods we employ as components of our proposing method.
In these methods, the goal is to build a detector $f$ for each semantic concept $c \in C$, where $C$ is a set of concepts.
We define notations to be used in this section and the next section as in Table~\ref{table1}.

\subsubsection*{Supervised Learning}
Supervised learning assumes that a set of training samples $\mathcal{X} = \{(x_{i}, y_{i})\}_{i=1}^{N}$ is given for each concept $c \in C$, where
$x_{i} \in \mathbb{R}^{d}$ is a feature vector of an image or video,
$y_{i} \in \{+1, -1\}$ is a positive or negative label for $x_{i}$,
and $N$ is the number of training samples.
A supervised detector $f_{\mbox{\tiny SV}}$ is then trained from these samples.
Its simplest example is a linear detector given by
\begin{align}
 \label{sv}
 f_{\mbox{\tiny SV}}(x) = \sum_{i=1}^{N} \alpha_{i} x_{i}^{T} x + \gamma,
\end{align}
where $x$ is a testing sample, and $\alpha_{i}$ and $\gamma$ are the model parameters.
Recent studies extend it to non-linear detectors by introducing kernel tricks \cite{Hofmann_2008, Aizerman_1964} and/or deep neural networks \cite{Krizhevsky_2012, Szegedy_2015}
.
Note that, by introducing explicit feature maps \cite{Vedaldi_2011} corresponding to kernel functions 
or by viewing deep neural networks as a feature extractor, these non-linear detectors often can be re-formulated as linear detectors in a high-dimensional feature space.

\tableA

\subsubsection*{Zero-Shot Learning}
Zero-shot learning assumes that a set of pre-trained detectors $\mathcal{P} = \{g_{j}(\cdot)\}_{j=1}^{M}$ is given instead of a set of training samples.
Here, $g_{j}$ is a pre-trained detector for a concept $d_{j} \in D$, where $D$ is another set of concepts disjoint to $C$,
and $M$ is the number of pre-trained detectors.
To build a detector for a concept $c \in C$ ($c \not \in D$),
recent zero-shot learning methods \cite{Norouzi_2014, Jain_2015, Xian_2017} combine given detectors by
\begin{align}
 \label{zs}
 f_{\mbox{\tiny ZS}}(x) = \sum_{j=1}^{M} \beta_{j} g_{j}(x) + \gamma',
\end{align}
where $\beta_{j}$ and $\gamma'$ are weighting and bias parameters, respectively.
Since $\beta_{j}$ is a weight which relates the concept $d_{j}$ to the target concept $c$,
the similarity measure between $d_{j}$ and $c$, $\mbox{sim}(d_{j},c)$,
is often used as $\beta_{j}$, i.e., $\beta_{j} = \mbox{sim}(d_{j},c)$.
Its example is cosine similarity between word vectors \cite{Mikolov_2013_2} given by
\begin{align}
\label{eqsim}
\mbox{sim}(d_{j},c) = \frac{\psi(d_{j})^{T} \psi(c)}{\|\psi(d_{j})\|_{2} \|\psi(c)\|_{2}},
\end{align}
where $\psi(\cdot) \in \mathbb{R}^{d'}$ is a word vector of a concept.
A word vector, which is a word representation by a real-valued vector, is obtained from semantic embedding methods such as skip-gram \cite{Mikolov_2013_2}.
The bias parameter $\gamma'$ is often experimentally optimized.

\section{Few-Shot Adaptation}
\label{sec:proposedmethod}

\subsection{Overview}
Our basic idea of few-shot adaptation is to optimize the parameters of supervised learning and zero-shot learning, simultaneously.
As shown in Figure~\ref{fig001}, the proposed framework accepts a pair of the following two sets as inputs:
\vspace{-0.1cm}
\begin{enumerate}
\setlength{\leftskip}{0.3cm}
\setlength{\itemsep}{0cm}
\item[(1)] a set of training samples $\mathcal{X}$,
\item[(2)] a set of pre-trained detectors $\mathcal{P}$,
\end{enumerate}
\vspace{-0.1cm}
which are from supervised and zero-shot learning frameworks, respectively.

To bridge supervised learning and zero-shot learning,
we impose the following two constraints: 
\begin{enumerate}
\setlength{\leftskip}{0.3cm}
\item[(C1)] few-shot adaptation outputs a supervised detector $f_{\mbox{\tiny SV}}$ if the set of pre-trained detectors is empty ($\mathcal{P} = \emptyset$).
\item[(C2)] few-shot adaptation outputs a zero-shot detector $f_{\mbox{\tiny ZS}}$ if the set of training samples is empty ($\mathcal{X} = \emptyset$).
\end{enumerate}

To simultaneously optimize the parameters of supervised learning and zero-shot learning,
few-shot adaptation linearly combines a supervised detector $f_{\mbox{\tiny SV}}$ and a zero-shot detector $f_{\mbox{\tiny ZS}}$,
i.e., we define a detector in few-shot adaptation by
\begin{align}
f_{\mbox{\tiny FS}}(x) = f_{\mbox{\tiny SV}}(x) + f_{\mbox{\tiny ZS}}(x).
\end{align}
For example, with a linear supervised detector in Eq.~(\ref{sv}) and a zero-shot detector in Eq.~(\ref{zs}),
we have
\begin{align}
\label{eqfs}
f_{\mbox{\tiny FS}}(x)
= \sum_{i=1}^{N} \alpha_{i} x_{i}^{T} x + \sum_{j=1}^{M} \beta_{j} g_{j} (x) + \gamma''.
\end{align}
The goal is to optimize $\alpha_{i}, \beta_{j},$ and $\gamma''$, where the two bias parameters are unified into $\gamma'' = \gamma + \gamma'$.

\tableB

We believe that this is a straightforward way to unify two learning frameworks, and expect that few-shot adaptation will be effective in cases where the number of training samples is small, because zero-shot learning and supervised learning benefit from each other.

\subsection{Introducing an Objective Function from Supervised Learning}

To satisfy the constraint (C1), an objective function should be imported from supervised learning to optimize the parameters in Eq.~(\ref{eqfs}).
However, since supervised learning is a mapping from a set of training samples $\mathcal{X}$ to a detector $f_{\mbox{\tiny SV}}$,
it can not be directly applied to the input $(\mathcal{X}, \mathcal{P})$ of few-shot adaptation, as inputs and outputs are summarized in Table~\ref{table2}.

To solve this problem, our idea is to generate a set of {\it pseudo} training samples $\mathcal{X}_{\mathcal{P}}$ from pre-trained detectors, and to apply supervised learning to a union set $\mathcal{U} = \mathcal{X} \cup \mathcal{X}_{\mathcal{P}}$.
In this way, many types of supervised learning techniques can be introduced to our framework without modifying their objective function.
Note that by simply defining $\mathcal{X}_{\emptyset} = \emptyset$,
we have $\mathcal{U} = \mathcal{X} \cup \mathcal{X}_{\emptyset} = \mathcal{X}$ when $\mathcal{P} = \emptyset$.
This shows that the constraint (C1) is satisfied.
The definition of
$\mathcal{X}_{\mathcal{P}}$
is given in the following subsection.

\subsection{Generating Pseudo Training Samples from Zero-Shot Detectors}

Our next focus is on the constraint (C2) for a zero-shot detector.
If $\mathcal{X} = \emptyset$, few-shot adaptation applies a supervised learning method to a set $\mathcal{U} = \mathcal{X} \cup \mathcal{X}_{\mathcal{P}} = \mathcal{X}_{\mathcal{P}}$ as described above.
We focus on {\it how to generate pseudo training samples that give a zero-shot detector as a result of supervised learning}.

Let us start from the simplest example using linear function
as the supervised detector and the zero-shot detector in Eq.~(\ref{eqfs}).
Let the pre-trained detectors for zero-shot learning be given by $g_{j}(x) = w_{j}^{T} x$.
Then,
\begin{align}
\label{eqlin}
f_{\mbox{\tiny FS}}(x)
= \sum_{i=1}^{N} \alpha_{i} x_{i}^{T} x + \sum_{j=1}^{M} \beta_{j} w_{j}^{T} x + \gamma''.
\end{align}
Here, $w_{j}$ ($\|w_{j}\|=1$) is the normal vector to the decision boundary of $g_{j}(x) = 0$.
In this equation, we see that its two terms on the right-hand side share a common structure that 
each term is a product of a parameter ($\alpha_{i}$ and $\beta_{j}$)
and an inner product of two vectors ($x_{i}^{T} x$ and $w_{j}^{T} x$).
Since $x_{i}$ can be understood as a training sample to optimize $\alpha_{i}$,
this one-to-one correspondence implies to us that $w_{j}$ can be used as a pseudo training sample to optimize $\beta_{j}$.

How to make pseudo training samples?
To obtain a function $g(x) = w^{T} x$ as a result of supervised learning,
the easiest way is to have a pair of the normal vector $w$ and its mirrored vector $-w$ for training with positive and negative labels, respectively, as shown in Figure~\ref{fig002} (b).
In this case, a set of pseudo training samples is given by $\mathcal{X}_{\mathcal{P}} = \{(\lambda w, +1),(-\lambda w, -1)\}$,
where $\lambda > 0$ is a scaling coefficient.

\figw{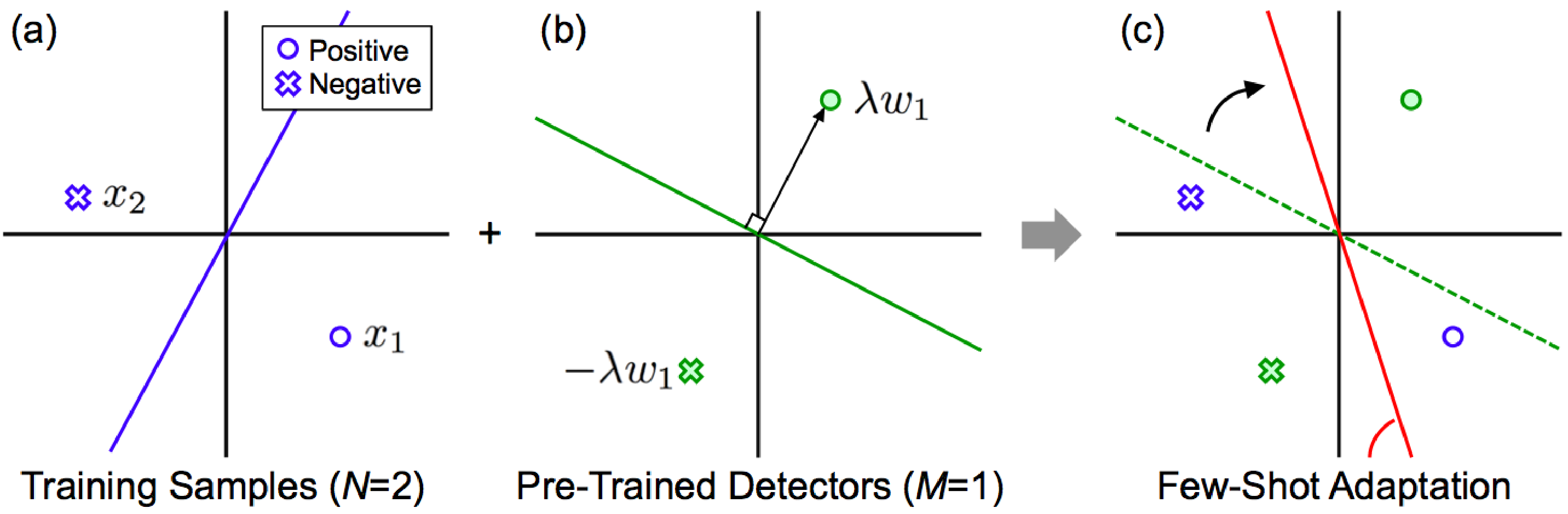}{1.0}{{\bf Example of few-shot adaptation.} (a) Training samples from a supervised learning framework. $x_{1}$ and $x_{2}$ are positive and negative training samples, respectively. (b) Pre-trained detectors from zero-shot learning framework. An example using a single pre-trained detector $g_{1}(x) = w_{1}^{T} x$ is given with illustration of the normal vector $\lambda w_{1}$. (c) Few-shot adaptation with a decision boundary, in which supervised learning and zero-shot learning are combined.}

This can be extended to a zero-shot detector $f_{\mbox{\tiny ZS}}$, a weighted sum of functions $g_{j}$.
By multiplying weight values $\mbox{sim}(d_{j},c)$ given from the zero-shot learning framework, e.g., Eq.~(\ref{eqsim}),
a set $\mathcal{X}_{\mathcal{P}} = \{( {\tilde x}_{k}, {\tilde y}_{k} )\}_{k=1}^{K}$ is defined by
\begin{align}
\label{pseudo1}
{\tilde x}_{2j} = + \lambda \mbox{sim} (d_{j},c) w_{j},\\
\label{pseudo2}
{\tilde x}_{2j-1} = - \lambda \mbox{sim} (d_{j},c) w_{j},
\end{align}
with ${\tilde y}_{2j} = +1, {\tilde y}_{2j-1} = -1$ for $j=1, 2, \cdots, M$, where $K = 2M$ is the number of pseudo training samples.

Finally, by applying supervised learning to a union set  $\mathcal{U} = \mathcal{X} \cup \mathcal{X}_{\mathcal{P}}$,
the detector of few-shot adaptation is reformulated by
\begin{align}
\label{eqfinal}
f_{\mbox{\tiny FS}}(x) = \sum_{i=1}^{N} \alpha_{i} x_{i}^{T} x + \sum_{k=1}^{K} {\tilde \beta}_{k} {\tilde x}_{k}^{T} x + \gamma'',
\end{align}
where $\alpha_{i}, {\tilde \beta}_{k}$, and $\gamma''$ are parameters.
Figure~\ref{fig002} shows how few-shot adaptation works with the minimum sets of training data.

To exactly obtain a zero-shot detector, the objective function imported from supervised learning for parameter estimation is required to give ${\tilde \beta}_{2j} = +1$ and ${\tilde \beta}_{2j-1} = -1$ when $\mathcal{X}=\emptyset$.
In practice, this is often a trivial solution of parameter estimation
since $\mathcal{U} = \emptyset \cup \mathcal{X}_{\mathcal{P}} = \mathcal{X}_{\mathcal{P}}$ only has pairs of symmetric samples.
In this case, we have
\begin{align}
f_{\mbox{\tiny FS}}(x)
&= \sum_{k=1}^{K} {\tilde \beta}_{k} {\tilde x}_{k}^{T} x + \gamma''\\
&= \sum_{j=1}^{M} ( {\tilde \beta}_{2j} {\tilde x}_{2j} + {\tilde \beta}_{2j-1} {\tilde x}_{2j-1} )^{T} x + \gamma''\\
&= \sum_{j=1}^{M} \lambda ( {\tilde \beta}_{2j} - {\tilde \beta}_{2j-1}) \mbox{sim} (d_{j},c) w_{j}^{T} x + \gamma''\\
&= 2 \lambda f_{\mbox{\tiny ZS}} (x),
\end{align}
and thus by setting $\lambda = \frac{1}{2}$, $f_{\mbox{\tiny FS}}$ is equal to $f_{\mbox{\tiny ZS}}$ as required in (C2).
Note also that if the dimension of feature vector $x$ is larger than the number of pre-trained detectors $M$,
$\mathcal{X}_{\mathcal{P}}$ becomes linearly separable.
This supports many supervised learning methods to satisfy the requirement,
by introducing recent high-dimensional feature extractor including deep convolutional networks.

\subsection{Extensions to Non-Linear Functions}
This subsection presents three methods to introduce nonlinearity to few-shot adaptation.
The first method extends linear few-shot adaptation in Eq.~(\ref{eqfinal}) to kernelized few-shot adaptation.
The second method introduces a deep convolutional network to a zero-shot detector in our framework.
The third method extends our framework to multi-class classification using neural networks.

\subsubsection{Kernelized Few-Shot Adaptation}

To introduce non-linearity into our framework, we apply kernel tricks \cite{Hofmann_2008,Aizerman_1964} to Eq.~(\ref{eqfinal}),
by replacing dot products with a kernel $\kappa(\cdot,\cdot)$.
The detector is then given by
\begin{align}
\label{fskernel}
f_{\mbox{\tiny FS}}(x) = \sum_{i=1}^{N} \alpha_{i} \kappa (x_{i}, x) + \sum_{k=1}^{K} {\tilde \beta}_{k} \kappa ({\tilde x}_{k}, x) + \gamma''.
\end{align}
Note that we keep to use a set $\mathcal{U} = \mathcal{X} \cup \mathcal{X}_{\mathcal{P}}$ for training.
With this kernelization, a linear zero-shot detector $f_{\mbox{\tiny ZS}}$ will not be obtained exactly when $\mathcal{X} = \emptyset$.
Instead, it provides a kernelized zero-shot detector, which can be viewed as an extended method for zero-shot learning.

\subsubsection{Pre-trained Detectors with Neural Networks}

In practice, many recent studies have proved the effectiveness of neural networks trained on large-scale datasets.
Most of these networks have a softmax classifier at the final layer.
Here, we present a way to introduce them to our framework by defining $w_{j}$ in Eq.~(\ref{pseudo1}) and (\ref{pseudo2}) by a concatenation of network parameters.

Let $h$ be an input of a softmax layer, which has outputs for $M$ concepts.
The output value (posterior probability) for the $j$-th concept is given by
\begin{align}
\label{softmax}
p_{j}(h) = \frac{\exp(a_{j}^{T} h + b_{j})}{\sum_{j=1}^{M} \exp(a_{j}^{T} h + b_{j})}
\end{align}
where $a_{j}$ and $b_{j}$ are parameters on the layer.
To extract pseudo training samples, we focus on its linear calculation $a_{j}^{T} h + b_{j}$, and define
\begin{align}
w_{j} = \left(
\begin{array}{c}
a_{j}\\
b_{j}
\end{array}
\right),
\end{align}
with a feature vector
\begin{align}
x = \left(
\begin{array}{c}
h\\
1
\end{array}
\right).
\end{align}
Note that $\exp$ function and the normalization process (with a denominator of the sum of $\exp$ values) in Eq.~(\ref{softmax}) are omitted with these definition.
They can be again introduced by utilizing Gaussian kernel in kernelized few-shot adaptation in Eq.~(\ref{fskernel}),
and by applying score normalization to values of $f_{\mbox{\tiny FS}}(x)$ if they are needed.


\subsubsection{Extension to Multi-Class Classification}
Our framework presented above is for binary classification to train concept detectors in a one-versus-all manner.
This is effective to detection tasks in the wild, e.g., TRECVID Semantic Indexing Task \cite{2015trecvidover}, in which a video shot can have multiple labels.
On the other hand, a number of recent studies on few-shot learning \cite{Vinyals_2016,Wang_2016,Hariharan_2017} have focused on multi-class classification using neural networks, for exmaple, object recognition on the ImageNet dataset.
To compare our approach with them, we extend our framework to multi-class classification by utilizing only positive pseudo training samples, i.e., a set of positive pseudo training samples $\mathcal{X}^{c}_{\mathcal{P}}$ for each concept $c$ is added to training samples for multi-class classification.


\figw{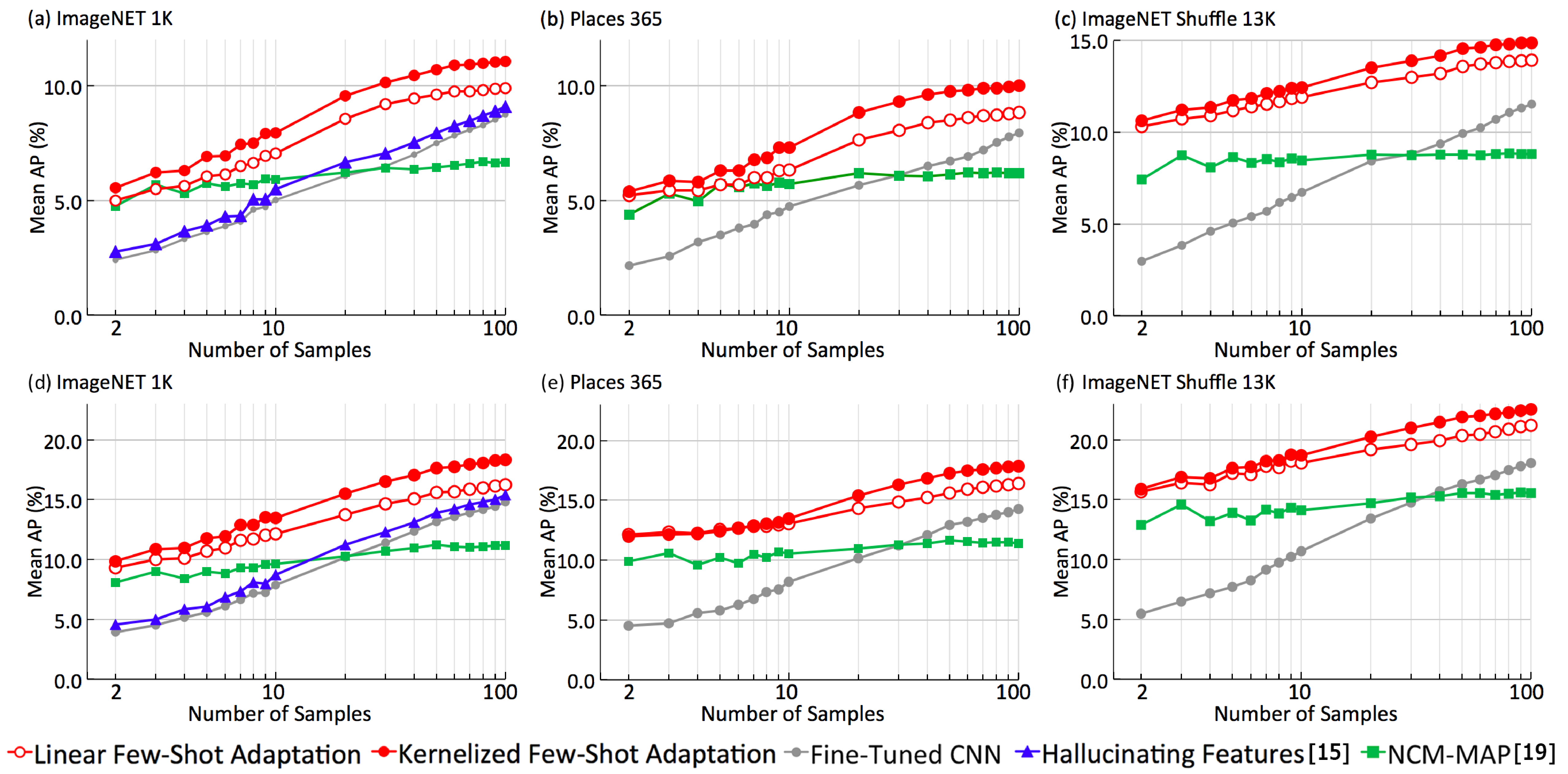}{1}{{\bf Few-Shot Evaluation on the TRECVID 2010 and 2014 datasets.}
Linear/Kernelized Few-Shot Adaptation: our proposed framework.
Fine-Tuned CNN: baseline using GoogLeNet features.
Hallucinating Features \cite{Hariharan_2017}: artificially-generated samples using neural networks trained using analogy among ImageNet objects.
NCM-MAP \cite{Mensink_2013}: Nearest class mean classifier with Maximum a Posteriori estimation using zero-shot priors.
Results are reported for three types of pre-trained networks: (a,d) ImageNet-1K, (b,e) Places-365, and (c,f) ImageNet-Shuffle13K,
on two datasets: (a,b,c) TRECVID 2010 and (d,e,f) TRECVID 2014.
All experiments are repeated for 10 times, and the average results are reported.
}

\section{Experiments}
\label{sec:experiments}
Our few-shot adaptation framework is evaluated on three datasets, TRECVID 2010, TRECVID 2014, and ImageNet.
\subsection{Evaluation on TRECVID datasets}
\subsubsection{Experimental Settings}

The TRECVID 2010 and 2014 datasets consist of Internet videos used in the TRECVID Semantic Indexing Competition \cite{2015trecvidover}.
Here, we use the whole ImageNet images \cite{Russakovsky_2015} and Places 365 images \cite{Zhou_2017} for pre-training.
We believe this is one of the best choices to report results by increasing the number of training samples from zero to many,
and to show the versatility of ImageNet and Places 365 datasets with our proposed framework.

The task is to detect semantic concepts from each video shot.
Shot boundaries are provided in the datasets.
The number of video shots for training and testing are listed in Table~\ref{table:setting}.
Each dataset has 30 types of semantic concepts to be detected.
The evaluation measure is Mean Average Precision (Mean AP),
which is calculated by using the official toolkit and annotations.

Evaluation results are reported on three training conditions: zero-shot, few-shot, and many-shot.
The zero-shot condition does not use TRECVID videos for training.
The few-shot condition limits the number of training video shots to $N$ ($0 < N \leq 100$) by random sampling.
The many-shot condition use all TRECVID videos for training.
In the many-shot condition, we can compare our results with official submissions using supervised learning methods at the competition.
Note that our main focus is on the few-shot condition.

For pre-trained detectors, three types of GoogLeNets \cite{Szegedy_2015} are used: ImageNet-1K, Places-365, and ImageNet-Shuffle13K.
The Goog-LeNet is a convolutional neural network with 23 layers.
ImageNet-1K uses 1.2 million images of 1,000 objects in ILSVRC 2012 for training \cite{Szegedy_2015}.
Places-365 uses 1.8 million images of 365 scenes in Places 2 dataset \cite{Zhou_2017}.
ImageNet-Shuffle13K uses the ImageNet Shuffle method \cite{Mettes_2016} for training, which provides 12,988 object classifiers trained on the whole ImageNet dataset.
For word vectors, 300 dimensional vectors obtained from the Skip-gram model in \cite{Mikolov_2013_2} are used.
They are used to measure similarity between concepts in Eq.~(\ref{eqsim}).
The average similarity between a TRECVID concept and its closest concept in pre-training is
0.557 for ImageNet-1K, 0.606 for Places-365 and 0.781 for ImageNet-Shuffle13K.
For objective function, the SVM loss (Hinge loss with $L_{2}$ regulalization) is used for parameter optimization.

\subsubsection{Few-Shot Evaluation}
Figure~\ref{fig003b} shows evaluation results on TRECVID 2010 and 2014 datasets under the few-shot condition.
As a baseline, results using CNN+SVM with supervised fine-tuning are reported, where the final softmax layer of GoogLeNet is replaced by an SVM.
Note that this has been one of the best ways to apply neural networks to the semantic indexing task at TRECVID.
For comparison, evaluation results of Nearest Class Mean Classifier \cite{Mensink_2013} with Maximum a Posteriori estimation using zero-shot priors (NCM-MAP) and artificially-generated hallucinating features (HF) in \cite{Hariharan_2017} are also reported.
HF is applied only for ImageNet-1K because it uses a network trained with analogy among the 1K objects on ILSVRC 2012.

We see from the results that kernelized few-shot adaptation performs the best with all networks and on both datasets.
This shows the effectiveness of the proposed framework,
and confirms that zero-shot learning and supervised learning benefit from each other when the number of training samples is small ($0 < N \leq 100$).
We also see that few-shot adaptation approaches to supervised fine-tuning as $N$ increases.
This shows our framework straightforwardly bridges zero-shot learning and supervised learning.

If we compare linear and kernelized few-shot adaptation, the kernelized one is always better.
This shows that the kernel trick in supervised learning is effective.
Utilizing the other types of kernels is interesting as a next step in future.

\tableC
\tableD
\tableE

\subsubsection{Zero-Shot Evaluation}
Table~\ref{table:zero-shot_evaluation} reports our results on the zero-shot condition.
Our method performs the best among methods in \cite{Jiang_2014b, McGuinness_2014} from the non-annotation track at TRECVID 2014 and zero-shot methods \cite{Norouzi_2014, Inoue_2016}.
Here, non-annotation track is a training condition which requires not to use TRECVID videos but to use the other resources such as web images for training.
Note that our zero-shot method can be viewed as a modification of ConSE \cite{Norouzi_2014},
where a normalization step for multi-class classification is omitted.
Since normalization assumes that each video shot has one of concept labels, it is not suitable for our concept detection task, in which a video shot can have multiple concept labels with unbalanced positive and negative samples.
To further improve the performance in the zero-shot condition, modifying and introducing recent zero-shot multi-class classification methods such as manifold learning \cite{Changpinyo_2016} is promising.

The results also show that late fusion of detection scores obtained from three networks improves the performance.
This suggests that adding other types of pre-trained detectors is also needed in future work.

\subsubsection{Many-Shot Evaluation}
Table~\ref{table:many-shot_evaluation} compares our results with the official submissions in TRECVID 2014.
We achieved 35.89 \% and 35.73\% in Mean AP with and without pseudo samples, respectively.
To the best of our knowledge, this is the best performance on this dataset.
This confirms that our pseudo training samples for few-shot adaptation do not affect the performance in the many-shot condition,
and means that our few-shot adaptation successfully unifies zero-shot learning and supervised learning.

\subsection{Evaluation on Imagenet dataset}

\subsubsection{Experimental Settings}
The ImageNet Large Scale Visual Recognition Competition (ILSVRC) dataset consists of 1.2 million images with 1,000 object categories.
For few-shot evaluation, we follow the evaluation setting proposed in \cite{Hariharan_2017},
which divides the 1,000 categories into 389 base categories and 611 novel categories.
All examples from base categories are used for pre-training, and few examples ($N = 1, 2, 5, 10,$ and $20$) for novel categories are used for few-shot adaptation.
Evaluation measure is Top-5 accuracy.
For a fair comparison, ResNet-10 used in \cite{Hariharan_2017} is applied to this multi-class classification problem.

\tableZ
\tableY

\figv{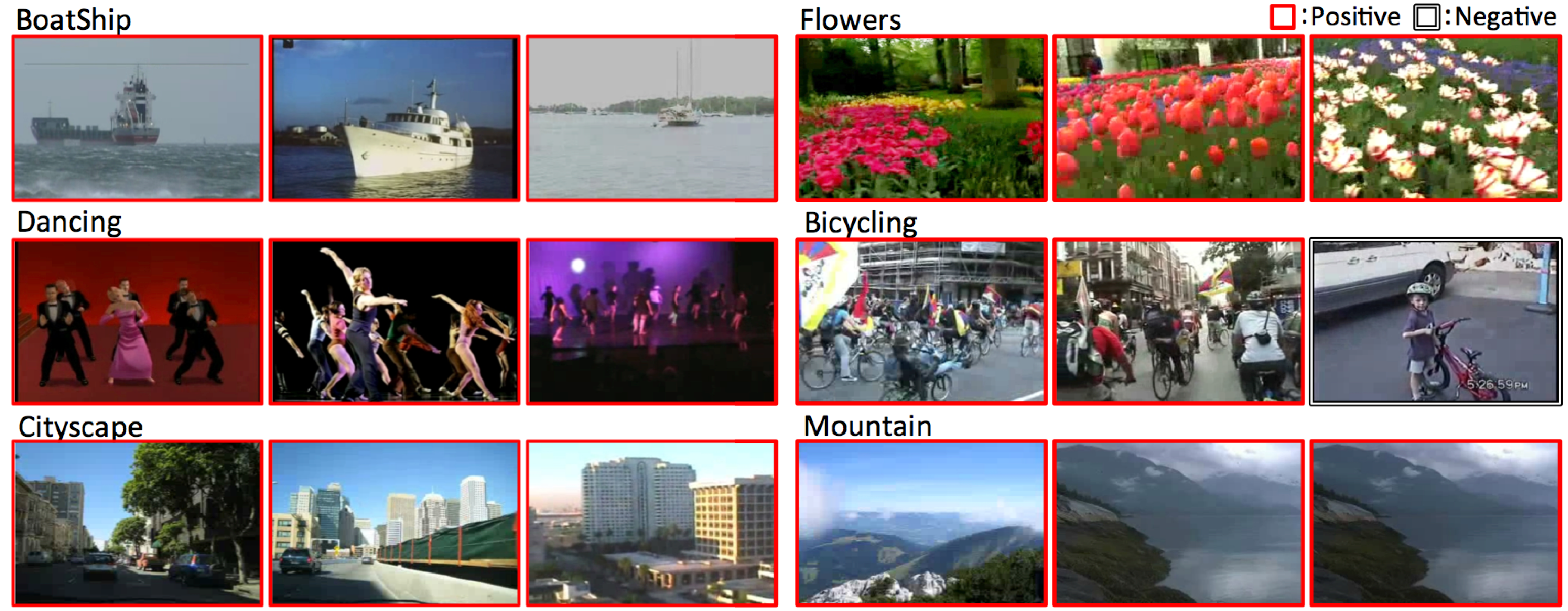}{1.0}{{\bf Top three video shots for six types of semantic concepts.} Kernelized few-shot adaptation with $N=10$ is applied.}
\figv{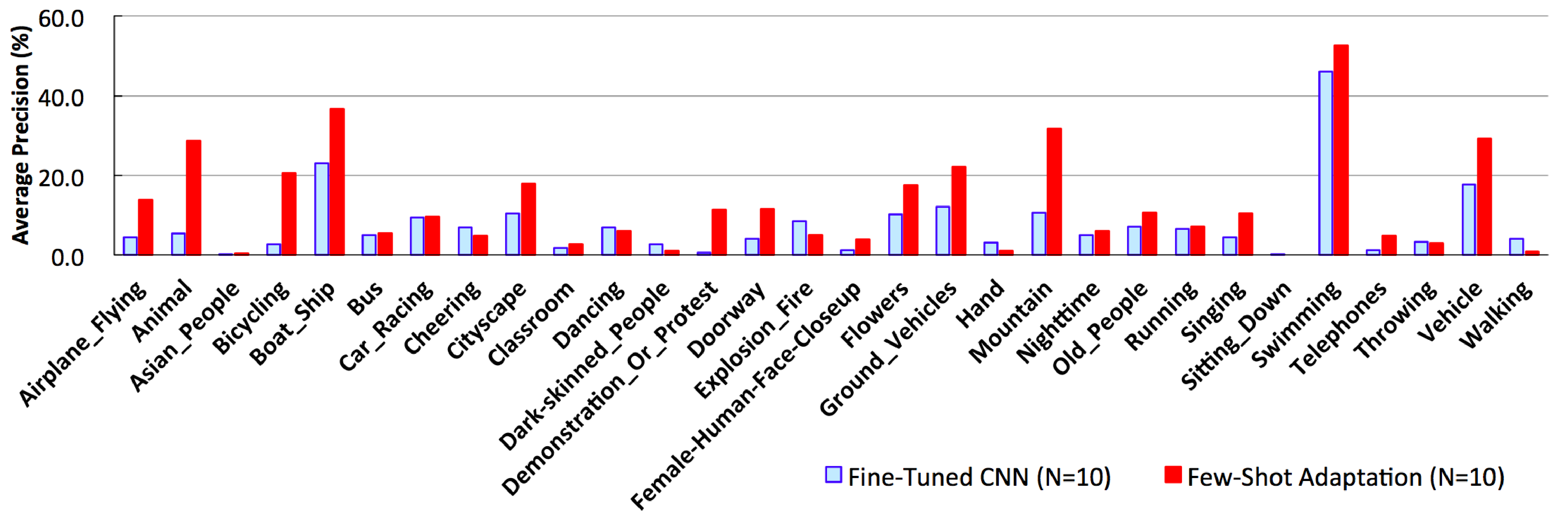}{1.0}{{\bf Average Precision by concepts on the TRECVID 2010 dataset.}}

\subsubsection{Comparison with Other Few-Shot Learning Methods}

Table~\ref{tableZ} compares our method with state-of-the-art few-shot learning methods for multi-class classification:
Matching Network (MN) for one-shot learning \cite{Vinyals_2016}, Model Regression \cite{Wang_2016}, and Hallucinating Features (HF) \cite{Hariharan_2017}.
We see our method performs the best for $N = 1,2,5,10$, and the second best for $N = 20$ among these methods.
To analyze results, Table~\ref{tableY} separately reports accuracy on novel and base sets of categories.
Note that they are in a trade-off relation.
We see a tendency that MN and HF are effective for novel and base categories, respectively, and that our method provides well-balanced performance on both.
This experiment uses only 389 pre-trained detectors for a fair comparison.
Increasing the number of base categories is promising to further improve the accuracy of our method.

\subsection{Limitations and Discussions}

Here we show some example video shots detected on the TRECVID dataset.
Figure~\ref{fig010} shows the top-ranked video shots.
They are mostly true positive shots even with 10 training examples.
Figure~\ref{fig004} shows Average Precisions (APs) by semantic concepts.
Few-shot adaptation is effective for various concepts.
However, it is difficult to detect actions such as Walking and SittingDown from video with few examples.
To further improve the overall performance, pre-trained detectors closely related to the domain of the TRECVID task are required.
For example, introducing 3D CNNs pre-trained on action video datasets such as Kinetics \cite{Kataoka_2018, Will_2018} would be interesting as a promising next step.


Another limitation of our framework is in its assumption that word vectors are given by a semantic embedding method.
New types of word vectors, such as those obtained by joint training of text and image representation \cite{Zhang_2017, Kodirov_2017} or manifold learning \cite{Changpinyo_2016}, can be introduced to our framework.
However, our few-shot adaptation can not update these embeddings in its training phase.
The three joint training of whole system including semantic embeddings is needed in future work.

\section{Conclusion}
\label{sec:conclusion}
We proposed a few-shot adaptation framework, which combines zero-shot learning and supervised learning.
It provided robust parameter estimation with few training examples,
by optimizing the parameters of zero-shot learning and supervised learning simultaneously.
Our experiments showed the effectiveness of the proposed framework on TRECVID 2010, 2014, and ImageNet datasets.
Our future work will be focusing on audio and text analysis to detect actions and events from video data.

\section*{Acknowledgement}
This work was supported by JST ACT-I Grant Number JPMJPR16U5 and JST CREST Grant Number JPMJCR1687, Japan. Part of this work is conducted as research activities of AIST-TokyoTech Real World Big-Data Computation Open Innovation Laboratory (RWBC-OIL).

\bibliographystyle{ACM-Reference-Format}

\bibliographystyle{ACM-Reference-Format}
\bibliography{sample-bibliography}

\begin{thebibliography}{99}
\bibitem{Snoek_2010}
C.~G.~M.~Snoek and M.~Worring.
Multimodal Video Indexing: A Review of the State-of-the-Art.
{\em In Springer Multimedia Tools and Applications}, vol.~25, no.~1, pp.~5--35, 2005.

\bibitem{Awad_2016}
G.~Awad, {\it et al.} 
TRECVid Semantic Indexing of Video: A 6-Year Retrospective.
{\em In ITE Trans. on Media Technology and Applications}, vol.~4, no.~3, pp.~187--208, 2016.

\bibitem{Zhao_2017}
B.~Zhao, {\it et al.} %
Hierarchical Recurrent Neural Network for Video Summarization.
{\em \proc  ACM Multimedia}, pp.~863--871, 2017.

\bibitem{Mei_2013}
T.~Mei, {\it et al.} 
Near-Lossless Semantic Video Summarization and Its Applications to Video Analysis.
{\em In ACM Trans. on Multimedia Computing, Communications, and Applications}, vol.~9, no.~3-16, pp.~1--23, 2013.

\bibitem{Xian_2017b}
Y.~Xian, {\it et al.} 
Evaluation of Low-Level Features for Real-World Surveillance Event Detection.
{\em In IEEE Trans. on Circuits and Systems for Video Technology}, vol.~27, no.~3, pp.~624--634, 2017.

\bibitem{Zhang_2014}
X.~Zhang, {\it et al.} 
Background-Modeling-Based Adaptive Prediction for Surveillance Video Coding.
{\em In IEEE Trans. on Image Processing}, vol.~23, no.~2, pp.~769--784, 2014.

\bibitem{Smeulders_2000}
A.~W.~M.~Smeulders, {\it et al.} 
Content-Based Image Retrieval at the End of the Early Years.
{\em In IEEE Trans. on PAMI}, vol.~22, no.~12, pp.~1349--1380, 2000.

\bibitem{Drucker_1997}
H.~Drucker, {\it et al.} 
Support Vector Regression Machines.
{\em \proc  NIPS}, pp. 155--161, 1997.

\bibitem{Vapnik_1995}
V.~Vapnik.
The Nature of Statistical Learning Theory. 
{\em Springer, New York}, 1995.

\bibitem{Krizhevsky_2012}
A.~Krizhevsky, {\it et al.} 
ImageNet Classification with Deep Convolutional Neural Networks.
{\em \proc  NIPS}, pp.1--9, 2012.

\bibitem{Simonyan_2015}
K.~Simonyan and A.~Zisserman.
Very Deep Convolutional Networks for Large-Scale Image Recognition.
{\em \proc  ICLR}, 2015.

\bibitem{Szegedy_2015}
C.~Szegedy, {\it et al.} 
Going Deeper with Convolutions.
{\em \proc  CVPR}, 2015.

\bibitem{Snoek_2014}
C.~G.~M. Snoek, {\it et al.} 
MediaMill at TRECVID 2014: Searching Concepts, Objects, Instances and Events in Video.
{\em \proc  TRECVID workshop}, 2014.

\bibitem{Perronnin_2010}
F.~Perronnin, {\it et al.} 
Improving the Fisher Kernel for Large-Scale Image Classification.
{\em \proc  ECCV}, 2010.

\bibitem{Hariharan_2017}
B.~Hariharan and R.~Girshick.
Low-shot Visual Recognition by Shrinking and Hallucinating Features.
{\em \proc  ICCV}, 2017.

\bibitem{Vinyals_2016}
O.~Vinyals, {\it et al.} 
Matching Networks for One Shot Learning.
{\em \proc  NIPS}, 2016.

\bibitem{Oquab_2014}
M.~Oquab, {\it et al.} 
Learning and Transferring Mid-Level Image Representations using Convolutional Neural Networks.
{\em \proc  CVPR}, 2014.

\bibitem{Babenko_2014}
A.~Babenko, {\it et al.} 
Neural Codes for Image Retrieval.
{\em \proc  ECCV}, pp.~584--599, 2014.


\bibitem{Mensink_2013}
T.~Mensink, J.~Verbeek, F.~Perronnin, and G.~Csurka.
Distance-Based Image Classification: Generalizing to New Classes at Near-Zero Cost.
{\em IEEE Trans. on PAMI}, vol.~35, no.~11, pp.~2624--2637, 2013.


\bibitem{Perronnin_Dance_Csurka_Bressan_2006}
F.~Perronnin, {\it et al.} 
\newblock Adapted Vocabularies for Generic Visual Categorization.
\newblock {\em \proc  ECCV}, pp.~464--475, 2006.

\bibitem{Norouzi_2014}
M.~Norouzi, {\it et al.} 
Zero-shot Learning by Convex Combination of Semantic Embeddings.
{\em \proc  ICLR}, 2014.

\bibitem{Frome_2013}
A.~Frome, {\it et al.} 
Devise: A Deep Visual-semantic Embedding Model.
{\em \proc  NIPS}, 2013.

\bibitem{Mensink_2014}
T.~Mensink, {\it et al.} 
Costa: Co-occurrence Statistics for Zero-shot Classification.
{\em \proc  CVPR}, 2014.

\bibitem{Jain_2015}
M.~Jain, {\it et al.} 
Objects2action: Classifying and Localizing Actions without Any Video Example.
{\em \proc  ICCV}, 2015.

\bibitem{Xian_2017}
Y.~Xian, {\it et al.} 
Zero-Shot Learning - the Good, the Bad and the Ugly.
{\em \proc  CVPR}, 2017.

\bibitem{Cappallo_2017}
S.~Cappallo and C.G.M.~Snoek.
Future-Supervised Retrieval of Unseen Queries for Live Video.
{\em \proc  ACM Multimedia}, pp.~28--36, 2017.

\bibitem{Qin_2017}
J.~Qin, {\it et al.} 
Zero-Shot Action Recognition With Error-Correcting Output Codes
{\em \proc  CVPR}, 2017.

\bibitem{Xu_2016}
X.~Xu, {\it et al.} 
Multi-Task Zero-Shot Action Recognition with Prioritised Data Augmentation.
{\em \proc  ECCV}, 2016.

\bibitem{Gan_2016}
C.~Gan, {\it et al.} 
Concepts not Alone: Exploring Pairwise Relationships for Zero-Shot Video Activity Recognition.
{\em \proc  AAAI}, pp.~3487--3493, 2016.

\bibitem{Mikolov_2013}
T.~Mikolov, {\it et al.} 
Efficient Estimation of Word Representations in Vector Space.
{\em \proc  ICLR}, 2013.

\bibitem{Mikolov_2013_2}
T.~Mikolov, {\it et al.} 
Distributed Representations of Words and Phrases and their Compositionality.
{\em \proc  NIPS}, 2013.

\bibitem{Gavves_2015}
E.~Gavves, {\it et al.} 
Active Transfer Learning with Zero-Shot Priors: Reusing Past Datasets for Future Tasks.
{\em \proc  ICCV}, 2015.

\bibitem{He_2016}
K.~He, {\it et al.} 
Deep Residual Learning for Image Recognition.
{\em \proc  CVPR}, 2016.

\bibitem{Huang_2017}
G.~Huang, {\it et al.} 
Densely Connected Convolutional Networks
{\em \proc  CVPR}, 2017.

\bibitem{Russakovsky_2015}
O.~Russakovsky, {\it et al.} 
ImageNet Large Scale Visual Recognition Challenge.
{\em In IJCV}, vol.115, no.3, pp.211--252, 2015.

\bibitem{Zhou_2017}
B.~Zhou, {\it et al.} 
Places: A 10 million Image Database for Scene Recognition.
{\em In IEEE Trans. on PAMI}, in press, 2017.



\bibitem{Santoro_2016}
A.~Santoro, {\it et al.} 
Meta-Learning with Memory Augmented Neural Network.
{\em \proc  ICML}, 2016.

\bibitem{Kwitt_2016}
R.~Kwitt, {\it et al.} 
One-Shot Learning of Scene Locations via Feature Trajectory Transfer.
{\em \proc  CVPR}, 2016.

\bibitem{Wang_2016}
Y.X.~Wang and M.~Hebert.
Learning to Learn: Model Regression Networks for Easy Small Sample Learning.
{\em \proc  ECCV}, 2016.

\bibitem{Inoue_Shinoda_2012}
N.~Inoue and K.~Shinoda.
\newblock A Fast and Accurate Video Semantic-Indexing System Using Fast MAP Adaptation and GMM Supervectors.
\newblock {\em IEEE Trans. on Multimedia}, vol.~14, no.~4, pp.~1196--1205, 2012.

\bibitem{Perronnin_Jorge_Mensink_2010}
F.~Perronnin, {\it et al.} 
\newblock Improving the fisher kernel for large-scale image classification.
\newblock {\em \proc  ECCV}, pp.~143--156, 2010.


\bibitem{Snoek_2015}
C.G.M.~Snoek, {\it et al.} 
Qualcomm Research and University of Amsterdam at TRECVID 2015: Recognizing Concepts, Objects, and Events in Video.
{\em \proc  TRECVID workshop}, 2015.

\bibitem{Mettes_2016}
P.~Mettes, {\it et al.} 
The ImageNet Shuffle: Reorganized Pre-training for Video Event Detection.
{\em \proc  ICMR}, 2016.

\bibitem{Habibian_2016}
A.~Habibian, {\it et al.} 
Video2vec Embeddings Recognize Events when Examples are Scarce.
{\em In IEEE Trans. on PAMI}, in press, 2017.

\bibitem{Feichtenhofer_2016}
C.~Feichtenhofer, {\it et al.} 
Convolutional Two-Stream Network Fusion for Video Action Recognition.
{\em \proc  CVPR}, 2016.

\bibitem{Sun_2015}
L.~Sun, {\it et al.} 
Human Action Recognition Using Factorized Spatio-Temporal Convolutional Networks.
{\em \proc  ICCV}, 2015.

\bibitem{Jiang_2014b}
L.~Jiang, {\it et al.} 
CMU-Informedia at TRECVID Semantic Indexing.
{\em \proc  TRECVID workshop}, 2014.

\bibitem{McGuinness_2014}
K.~McGuinness, {\it et al.} 
Insight Centre for Data Analytics at TRECVid 2014: Instance Search and Semantic Indexing.
{\em \proc  TRECVID workshop}, 2014.

\bibitem{Lampert_2013}
C.H.~Lampert, {\it et al.} 
Attribute-Based Classification for Zero-Shot Visual Object Categorization.
{\em In IEEE Trans. on PAMI}, vol.~36, no.~3, pp.~453--465, 2013.

\bibitem{Changpinyo_2016}
S.~Changpinyo, {\it et al.} 
Synthesized Classifiers for Zero-Shot Learning.
{\em \proc  CVPR}, 2016.

\bibitem{Huang_2015}
S.~Huang, {\it et al.} 
Learning Hypergraph-Regularized Attribute Predictors.
{\em \proc  CVPR}, 2015.

\bibitem{Yu_2010}
X.~Yu and Y.~Aloimonos.
Attribute-based Transfer Learning for Object Categorization with Zero or One Training Example.
{\em \proc  ECCV}, 2010.

\bibitem{Yu_2012}
Q.~Yu, {\it et al.} 
Multimedia event recounting with concept based representation.
{\em ACM Multimedia}, 2012.

\bibitem{Jiang_2014}
L.~Jiang, {\it et al.} 
Easy Samples First: Self-Paced Reranking for Zero-Example Multimedia Search.
{\em \proc  ACM Multimedia}, 2014.

\bibitem{Habibian_2014}
A.~Habibian, {\it et al.} 
Composite Concept Discovery for Zero-shot Video Event Detection.
{\em \proc  ICMR}, 2014.

\bibitem{Wu_Bondugula_Luisier_Zhuang_Natarajan_2014}
S.~Wu, {\it et al.} 
Zero-shot Event Detection using Multi-modal Fusion of Weakly Supervised Concepts.
{\em \proc  CVPR}, pp.2665--2672, 2014.

\bibitem{Inoue_2016}
N.~Inoue and K.~Shinoda.
Adaptation of Word Vectors using Tree Structure for Visual Semantics.
{\em \proc  ACM Multimedia}, pp.~277--281, 2016.

\bibitem{Zhang_2016}
Z.~Zhang and V.~Saligrama.
Zero-Shot Learning via Joint Latent Similarity Embedding.
{\em \proc  CVPR}, 2016.

\bibitem{Xian_2016}
Y.~Xian, {\it et al.} 
Latent Embeddings for Zero-shot Classification
{\em \proc  CVPR}, 2016.

\bibitem{Zhang_2017}
L.~Zhang, {\it et al.} 
Learning a Deep Embedding Model for Zero-Shot Learning.
{\em \proc  CVPR}, 2017.

\bibitem{Kodirov_2017}
E.~Kodirov, {\it et al.} 
Semantic Autoencoder for Zero-Shot Learning
{\em \proc  CVPR}, 2017.

\bibitem{Hofmann_2008}
T.~Hofmann, {\it et al.} 
Kernel Methods in Machine Learning.
{\em In The Annals of Statistics}, vol.~36, no.~3, pp.~1171--1220, 2008.

\bibitem{Aizerman_1964}
M.A.~Aizerman, {\it et al.} 
Theoretical Foundations of the Potential Function Method in Pattern Recognition Learning.
{\em In Automation and Remote Control}, vol.~25, pp.~821--837, 1964.

\bibitem{Vedaldi_2011}
A.~Vedaldi and A.~Zisserman.
Efficient Additive Kernels via Explicit Feature Maps.
{\em In IEEE Trans. on PAMI}, vol.~34, no.~3, pp.~480--492, 2012.


\bibitem{Jason_2014}
Y.~Jason, {\it et al.} 
How Transferable are Features in Deep Neural Networks?.
{\em \proc  NIPS}, pp. 3320--3328, 2014.

\bibitem{Shao_2015}
L.~Shao, {\it et al.} 
Transfer Learning for Visual Categorization: A Survey.
{\em In IEEE Trans. on Neural Networks and Learning Systems}, vol.~26, no.~5, pp.~1019--1034, 2015.

\bibitem{2015trecvidover}
G.~Awad, {\it et al.} 
TRECVID 2015 -- An Overview of the Goals, Tasks, Data, Evaluation Mechanisms and Metrics.
{\em \proc  TRECVID workshop}, 2015.

\bibitem{Will_2018}
W.~Kay, {\it et al.}
The Kinetics Human Action Video Dataset.
{\em arXiv preprint}, arXiv:1705.06950, 2017.

\bibitem{Kataoka_2018}
K.~Hara, {\it et al.}
Can Spatiotemporal 3D CNNs Retrace the History of 2D CNNs and ImageNet?.
{\em \proc  CVPR}, 2018.

\bibitem{Laaksonen_2014}
J.~Laaksonen, {\it et al.}
PicSOM Experiments in TRECVID 2014 Semantic Indexing Task.
{\em \proc  TRECVID workshop}, 2014.

\bibitem{Inoue_2014}
N.~Inoue and K.~Shinoda.
n-gram Models for Video Semantic Indexing.
{\em \proc  ACM Multimedia}, 2014.

\bibitem{Safadi_2015}
B.~Safadi, {\it et al.}
Descriptor Optimization for Multimedia Indexing and Retrieval.
{\em In Springer Multimedia Tools and Applications}, vol.~74, no.~4, pp.~1267--1290, 2015.

\bibitem{Ballas_2014}
N.~Ballas, {\it et al.}
Irim at TRECVID 2014: Semantic indexing and Instance Search.
{\em \proc  TRECVID workshop}, 2014.

\end{thebibliography}

\end{document}